\documentclass[pra,aps,showpacs,superscriptaddress,reprint]{revtex4-1}

\usepackage{amsmath}
\usepackage{graphicx}
\graphicspath{{figures/}}
\usepackage{epsfig}
\usepackage{float}
\usepackage{physics}
\usepackage{tikz}
\usetikzlibrary{calc, intersections, decorations.markings}
\usepackage{booktabs}

\def\ket#1{|#1\rangle}

\def\redmel#1#2#3{\langle#1\| #2 \| #3 \rangle}
\def\eval#1{\left\langle #1 \right\rangle}

\def\ms{\,\mathrm{ms}}
\def\us{\,\mathrm{\micro s}}

\def\nT{\,\mathrm{nT}}
\def\nm{\,\mathrm{nm}}

\def\kHz{\,\mathrm{kHz}}
\def\Hz{\,\mathrm{Hz}}
\def\MHz{\,\mathrm{MHz}}

\def\us{\,\mu\mathrm{s}}

\def\eref#1{Eq.~(\ref{#1})}
\def\fref#1{Fig.~\ref{#1}}

\begin{document}
\title{Hyperfine-mediated effects in a Lu$^+$ optical clock}
\author{Zhang Zhiqiang}
\email{e0000155@u.nus.edu}
\affiliation{Centre for Quantum Technologies, 3 Science Drive 2, 117543 Singapore}
\author{K. J. Arnold}
\affiliation{Centre for Quantum Technologies, 3 Science Drive 2, 117543 Singapore}
\affiliation{Temasek Laboratories, National University of Singapore, 5A Engineering Drive 1, 117411 Singapore}
\author{R. Kaewuam}
\affiliation{Centre for Quantum Technologies, 3 Science Drive 2, 117543 Singapore}
\author{M. S. Safronova}
\affiliation{Department of Physics and Astronomy, University of Delaware, Newark, Delaware 19716, USA}
\affiliation{Joint Quantum Institute, National Institute of Standards and Technology and the University of Maryland,
College Park, Maryland, 20742}
\author{M. D. Barrett}
\email{phybmd@nus.edu.sg}
\affiliation{Centre for Quantum Technologies, 3 Science Drive 2, 117543 Singapore}
\affiliation{ Department of Physics,National University of Singapore,  2 Science Drive 3, 117551 Singapore}
\begin{abstract}
We consider hyperfine-mediated effects for clock transitions in $^{176}$Lu$^+$. Mixing of fine structure levels due to the hyperfine interaction bring about modifications to Land\'e $g$-factors and the quadrupole moment for a given state.  Explicit expressions are derived for both $g$-factor and quadrupole corrections, for which leading order terms arise from the nuclear magnetic dipole coupling.  High accuracy measurements of the $g$-factors for the $^1S_0$  and $^3D_1$ hyperfine levels are carried out, which provide an experimental determination of the leading order correction terms.
\end{abstract}
\maketitle
\section{Introduction}
Singly ionized lutetium ($^{176}$Lu$^+$) is a unique optical clock candidate in that it provides three possible clock transitions. Of particular interest in this work is the $^1S_0\leftrightarrow{}^3D_1$ transition at 848\,nm, which has favourable clock properties relative to leading clock candidates \cite{arnold2018blackbody}.  At the Doppler cooling limit, the $^3D_1\leftrightarrow{}^3P_0$ cooling transition provides a fractional second-order Doppler shift below $10^{-19}$.  The large atomic mass and additional clock transitions allow micromotion shifts to be controlled to a similar level.  The blackbody radiation (BBR) shift of the 848-nm transition is $-1.36(10)\times 10^{-18}$ at 300\,K, which is the lowest of any optical clock system \cite{arnold2018blackbody} and easily controllable to the low 10$^{-19}$ with modest technical effort.  More recently, experiments have demonstrated the potential for clock operation with multiple ions, which will ultimately provide improved stability \cite{tan2019suppressing,kaewuam2020hyperfine}.  Thus it can be anticipated that this transition will ultimately provide an error budget competitive with leading systems.

A crucial consideration for clock implementation with $^{176}$Lu$^+$ is the use of hyperfine averaging in which a reference frequency is defined by an average over all hyperfine states with a common magnetic quantum number, $m$ \cite{barrett2015developing}.   Provided $|m|<I-J$, where $I$ is the nuclear spin and $J$ is the electronic angular momentum, the averaging realizes an effective $J=0$ level and practically eliminates dominant Zeeman shifts and shifts arising from rank 2 tensor interactions, such as the electric quadrupole moment \cite{barrett2015developing}.  The averaging principle holds even when there is a large amount of Zeeman mixing within a given fine-structure level, but it omits hyperfine-mediated mixing with other levels.  Such mixing influences $g$-factors \cite{porsev2017theoretical} and is the mechanism for the non-zero quadrupole moment of $^3P_0$ clock states in Al$^+$ and In$^+$ \cite{beloy2017hyperfine}.  Consequently, it can be anticipated that similar effects will occur for $^{176}$Lu$^+$ and likely influence the effectiveness of hyperfine averaging.

In this paper, the influence of hyperfine-mediated mixing on the clock states of $^{176}$Lu$^+$ is investigated via high accuracy measurements of $g$-factors for the $^1S_0$  and $^3D_1$ hyperfine levels.  Comparison with theoretical results provides an experimental determination of the leading order correction terms, which arise from the nuclear magnetic dipole coupling.  As similar corrections also apply to the quadrupole moments of $^3D_1$ states, the measurements  also allow a reasonable estimate for the residual quadrupole moment arising from hyperfine averaging.  Although the corresponding shift of the clock frequency will likely be well below $10^{-18}$, it will inevitably be an important consideration for upcoming clock assessments for this atom.

\section{Experiment}
\subsection{Apparatus}
\begin{figure}[ht]
\includegraphics[width=0.5\textwidth]{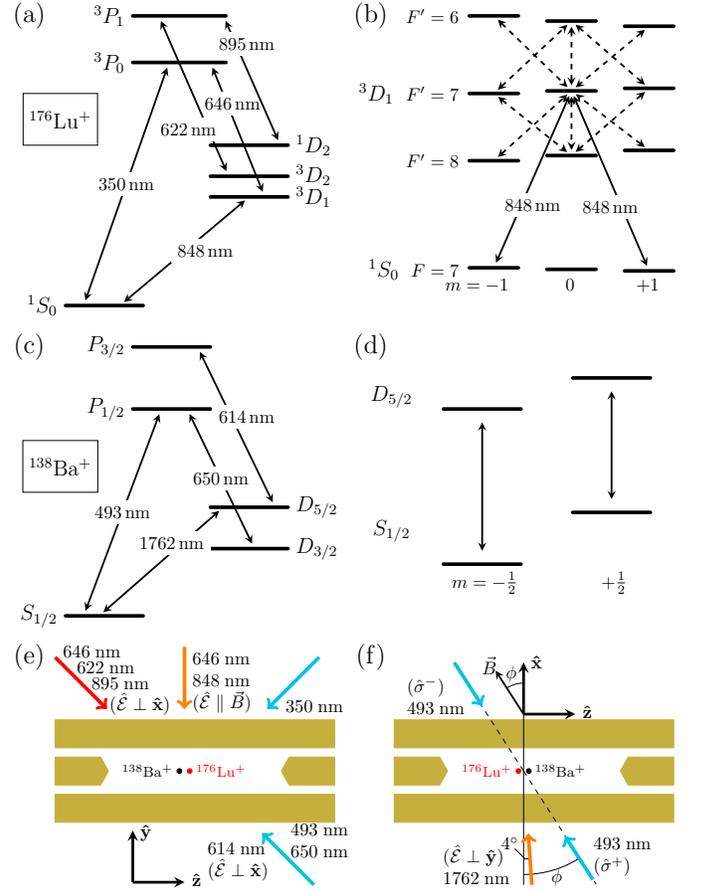}
\caption{Schematic of the experimental implementation. Level structures of (a) $^{176}$Lu$^+$  and  (c) $^{138}$Ba$^+$.  (b,d) Clock transitions used in this work from which Zeeman splittings are inferred. Dashed lines are microwave transitions. (e-f) Polarizations and geometric orientations of lasers.}
\label{fig:scheme}
\end{figure}

The relevant level structure of $^{176}$Lu$^+$ and the  laser systems required are shown in \fref{fig:scheme}(a-b). Lasers at $350\nm$, $895\nm$ and $622\nm$ provide optical pumping to the ${^3D_1}$ state. A laser at $646\nm$ provides Doppler cooling and state detection for the $^3D_1$ state with fluorescence collected onto either a single photon counting module (SPCM) or an EMCCD camera.  An additional $\pi$-polarized 646-nm laser addressing $F=7$ to $F' =7$ facilities state preparation into $\ket{{^3D}_1, 7, 0}$. A clock laser at $848\nm$ drives the $^1S_0 - {^3D_1}$ clock transition. Two microwave antennas are used to drive the $\Delta m =0,\pm1$ microwave transitions indicated in \fref{fig:scheme}(b).  On their respective microwave transitions, each antenna was positioned by hand to give approximately equal coupling to the $\Delta m = \pm 1$ transitions and reduced coupling for $\Delta m = 0$.

The relevant level structure of $^{138}$Ba$^+$ and the  laser systems required are shown in \fref{fig:scheme}(c-d). Doppler cooling is achieved by driving 493- and 650-nm transitions with fluorescence at 650\,nm collected for state detection. The $D_{5/2}$ level is populated by driving the clock transition at $1762\nm$ and and depopulated by optical pumping on the 614-nm transition. State preparation into $m = \pm\frac{1}{2}$ states of $S_{1/2}$ is provided by two additional $\sigma^{\pm}$ polarized 493-nm beams. 

The 848-nm clock laser is locked to a 10 cm long ultra-low expansion (ULE) cavity with finesse of $\sim 4\times10^5$ and has a line-width of $\sim$1 Hz. The 1762-nm laser is phased-locked to an optical frequency comb (OFC), which is itself phase-locked to the 848-nm laser. The short term stability ($\lesssim10\,\mathrm{s}$) of the OFC is thus derived from the ULE cavity. On longer time scales ($\gtrsim10 s$) the OFC is steered to an active hydrogen maser (HM) reference. All rf and microwave sources are referenced to the HM.

The configurations and polarizations of all laser beams relative to the trap are illustrated in Fig.~\ref{fig:scheme}(e)\&(f).  For reference purposes a coordinate system is given, where $\hat{\mathbf{x}}$ ($\hat{\mathbf{y}}$) horizontal (vertical) with respect to the table top and $\hat{\mathbf{z}}$ is along the trap axis.  The trap is a four-rod linear Paul trap with axial end caps as described in previous work \cite{arnold2020precision}. In this work, the trap drive frequency is $20.57\MHz$, and the measured trap frequencies for a single $^{138}$Ba$^+$ are $\sim 2\pi \times(912,795,226) \kHz$, with the lowest trap frequency along the trap axis.  As shown in \fref{fig:scheme}(f), a dc magnetic field is applied in the $xz$-plane at an angle $\phi = 33(2)^{\circ}$ with respect to $\hat{\mathbf{x}}$, which defines the quantization axis.

\subsection{Measurements}
The $g$-factors of the $^{176}$Lu$^+$ $^1S_0 (F=7)$ and ${^3D_1} (F'= 6,7,8)$ levels are denoted $g_I$ and $g_F'$ respectively, and are measured via a comparison of Zeeman splittings.  Comparisons between $^{176}$Lu$^+$ and $^{138}$Ba$^+$ enable determination of $r_6\equiv g_\mathrm{Ba}/g_6$ and $r_8\equiv g_\mathrm{Ba}/g_8$ where $g_{\mathrm{Ba}} \equiv \frac{1}{2}(g[D_{5/2}] - g[S_{1/2}])$. The $g$-factors for Lu$^+$ can then be inferred using the accurately known $g$-factors in $^{138}$Ba$^+$ \cite{marx1998precise,arnold2020precision}.  Ratios among $g_{F'}$ and $g_8/g_I$ are measured using a single ion.  Together the two sets of experiments provide a complete determination of $g_I$ and $g_{F'}$ as well as consistency checks between the measurements.

The ratios $r_6$ and $r_8$ are measured at an applied magnetic field of $\sim1.573\,$mT. The experiment sequence consists of the following steps: $200\us$ preparation of $^{176}$Lu$^+$ in $^3D_1$, Doppler cooling of $^{138}$Ba$^+$ and $^{176}$Lu$^+$ for $1 \ms$, $1 \ms$ of optical pumping $^{138}$Ba$^+$ to either $\ket{S_{1/2}, m = \pm\frac{1}{2}}$  and $^{176}$Lu$^+$ to $\ket{^3D_1,7,0}$, Rabi spectroscopy with a pulse duration of $1.5\ms$ being performed simultaneously on the $^{138}$Ba$^+$ $\ket{D_{3/2}, m = \pm\frac{1}{2}}$ transition and $^{176}$Lu$^+$ $\ket{^3D_1,7,0}$ to  $\ket{^3D_1,6,\pm{1}}$ (or $\ket{^3D_1,8,\pm{1}}$) transition, 8 ms shelving of the remaining  $^{176}$Lu$^+$ $\ket{^3D_1,7,0}$ population to $\ket{^1S_0,7,\pm1}$,  sequential state detection of $^{138}$Ba$^+$ and $^{176}$Lu$^+$ for $\lesssim$1 ms each, and finally preparation of $^{176}$Lu$^+$ in the $^3D_1$ state and detection for $20\ms$. The last step detects the position of $^{176}$Lu$^+$ in the two-ion crystal using the different photon collection efficiencies for the two possible crystal configurations. The sequence is repeated four times for Rabi interrogation of both full width half maximum of the respective pair of Zeeman transitions. Every 20 cycles an integrating servo is updated to track the respective Zeeman splittings for both $^{138}$Ba$^+$ and $^{176}$Lu$^+$.

\begin{figure}[ht]
\includegraphics[width=0.48\textwidth]{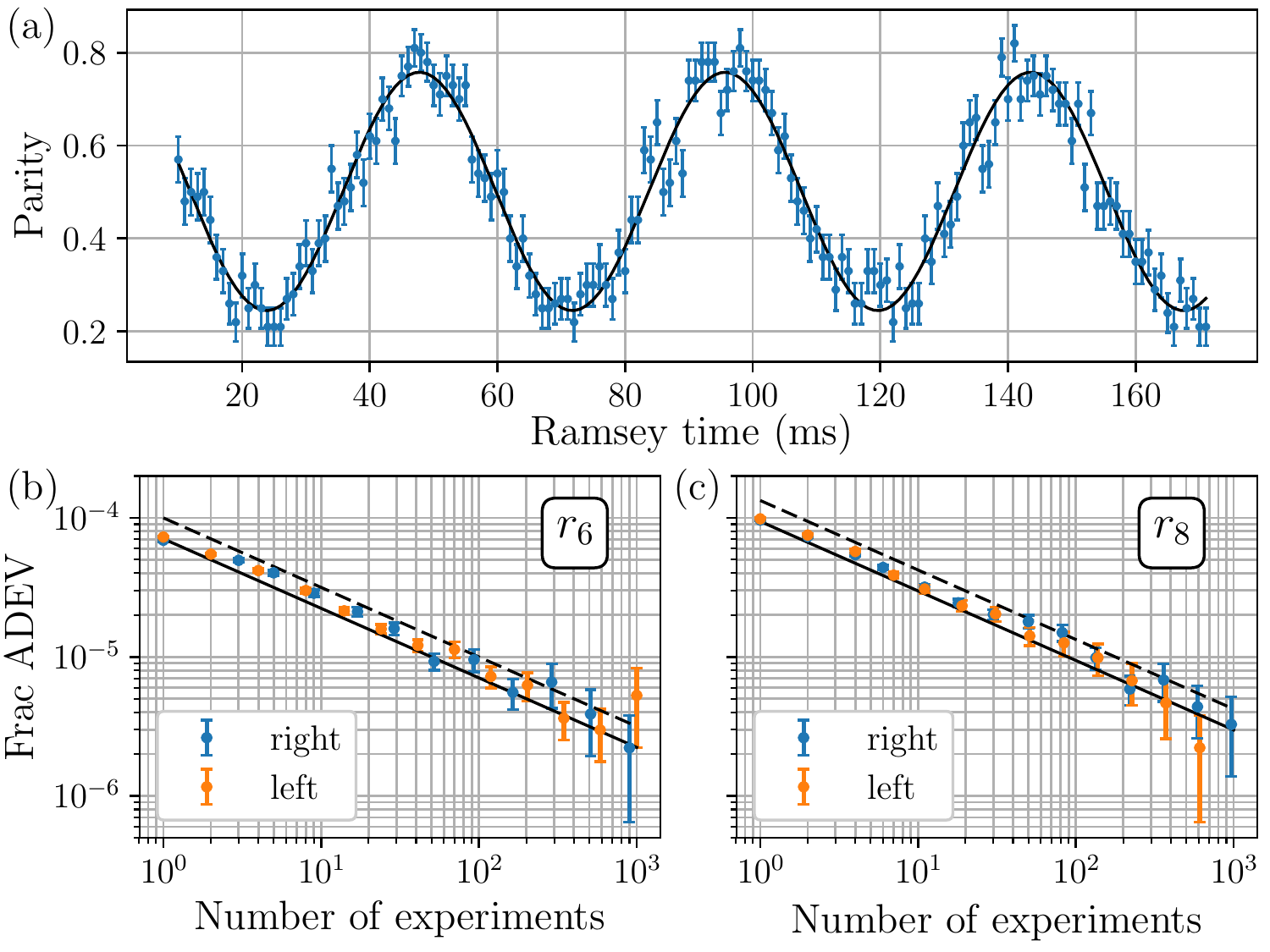}
\caption{(a) Correlation spectroscopy of the $\ket{S_{1/2},\pm\frac{1}{2}} - \ket{D_{3/2},\pm\frac{1}{2}}$ transition of two $^{138}$Ba$^+$ ions. The oscillation frequency of 20.92(8) Hz corresponds to a differential field of $3.73(1)\nT$ between the two ions.  (b)-(c) Fractional Allan deviation of $r_6$ and $r_8$ for Lu$^+$ on either the left (blue) or right (red) crystal position. The solid black is the projection noise limit. Dashed line is $\sqrt{2}$ above the projection noise.}
\label{fig:corr}
\end{figure}

To account for possible spatial dependence on the magnetic field, an additional experiment is performed to calibrate the gradient along the crystal axis. This is done using correlation spectroscopy \cite{chwalla2007precision,Quantumcoherence} on the $\ket{S_{1/2},\pm\frac{1}{2}} - \ket{D_{3/2},\pm\frac{1}{2}}$ transition in a two-ion crystal of $^{138}$Ba$^+$, similar to previous work \cite{tan2019suppressing}.  Specifically, Ramsey spectroscopy is performed on both ions for a duration longer than the optical coherence time of the individuals ions, which is limited by the common mode magnetic field noise. The EMCCD camera is used for single shot detection of both ions.  The parity, $p_{12} = \eval{\sigma_{z,1} \sigma_{z,2}}$, when averaged over all optical phases of the closing Ramsey pulse, is expected to yield $p_{12} = \frac{p_c}{2}\cos[2\pi(f_1 -f_2)T]$, where $p_c$ characterises the relative coherence between two oscillators, $f_{i}$ is the resonant frequency of the $i^{th}$ ion, and $T$ is the Ramsey time. Figure \ref{fig:corr}a shows the typical result as a function of Ramsey time.  The difference frequency between the ions measured before and after the measurements of $r_6$ and $r_8$ was found to be stable at $20.92(8)\Hz$, which corresponds to a magnetic field gradient of $0.3917(15)\, \mathrm{mT/m}$. 

The ratios of $g_{F'}$ are found by interleaved measurement of $\ket{^3D_1,F^\prime, \pm1}$ Zeeman splittings via microwave spectroscopy on a single Lu$^+$ ion with an applied magnetic field of $\sim 1.107\,$mT. The experimental sequence for measuring $g_6$ and $g_8$ is similar to measurements of $r_6$ and $r_8$ but without the Ba$^+$ lasers and a longer interrogation time of $16\ms$.  To measure $g_7$, additional microwave pulses to transfer from $\ket{^3D_1,7,0}$ to $\ket{^3D_1,6,0}$ or $\ket{^3D_1,8,0}$ are inserted as required. A single cycle consists of sequential Rabi interrogation of four Zeeman pairs: $\ket{^3D_1,6,\pm1}$, $\ket{^3D_1,8,\pm1}$, and $\ket{^3D_1,7,\pm1}$ twice, starting from either $\ket{^3D_1,6,0}$ or $\ket{^3D_1,8,0}$ to check for consistency. Every 20 cycles the four independent servos tracking the Zeeman splittings are updated. 

An additional experiment measures the ratio $g_{8}/{g_I}$ by interleaved measurement of the $\ket{^3D_1,8,\pm1}$ splitting using a $16\ms$  interrogation time on the microwave transition and the $\ket{^1S_0,7,\pm1}$ splitting using a $45\ms$ interrogation on the $848\nm$ optical transitions shown in \fref{fig:scheme}(b).  The $45\ms$ $\pi$-time allows for higher resolution of the much smaller ground state Zeeman splitting and ensures negligible probe induced shifts.

The $r_6$ and $r_8$ results are summarized in Table \ref{tab:baratio} with corrections given for the leading systematic effects. As the ions are observed to swap position every 18 minutes on average, which is much longer than the servo update period, data is sorted into the two possible crystal configurations.  Figure \ref{fig:corr}(b-c) shows the Allan deviations of $r_6$ and $r_8$ collected over the course of 5 hours. The Allan deviations are observed to be slightly elevated above the projection noise. We attribute this to the magnetic field noise which is comparable to the projection noise for the interrogation time used. To account for this, we take the statistical uncertainty in the resulting mean to be $\sqrt{2}$ above the projection noise limit, as indicated by the dashed black lines in \fref{fig:corr}(b-c). 

The leading systematic effects are from the magnetic field gradient, ac Stark shifts due to off-resonant microwave couplings in Lu$^+$, and the shift on the Ba$^+$ Zeeman splitting due to the ac magnetic fields at the trap rf frequency \cite{gan2018oscillating}.   Assessment of the magnetic field gradient has already been discussed, leaving only the shifts from the microwave and trap-induced ac fields.

To evaluate the ac Stark shifts from the microwave probe fields, the polarization components at the ion from each antenna were assessed from the relative coupling strength on $\Delta m = (-1,0,1)$ transitions at fixed rf power. For the 1.5 ms $\pi$-time used during the measurements of $r_6$ and $r_8$, we estimate the ac Stark shift to be $\pm0.21(2)\Hz$ on the $\ket{^3D_1,7,0} - \ket{^3D_1,6,\pm1}$ transitions and $\mp0.13(1)\Hz$ for the $\ket{^3D_1,7,0} - \ket{^3D_1,8,\pm1}$ transitions.  

Shifts from the trap-induced ac magnetic fields, depend only on the component of the ac field perpendicular to the applied dc field \cite{gan2018oscillating}.  This is measured from an Autler-Townes splitting exactly as described in previous work \cite{arnold2020precision}. The inferred field amplitude of $B_\perp = 1.25(1)\,\mu$T implies a  $\mp0.838(19)\Hz$ shift on the $\ket{S_{1/2},\pm\frac{1}{2}} - \ket{D_{3/2},\pm\frac{1}{2}}$ transitions at the operating magnetic field of $1.573(1)\,$mT. 

\begin{figure}[ht]
\includegraphics[width=0.48\textwidth]{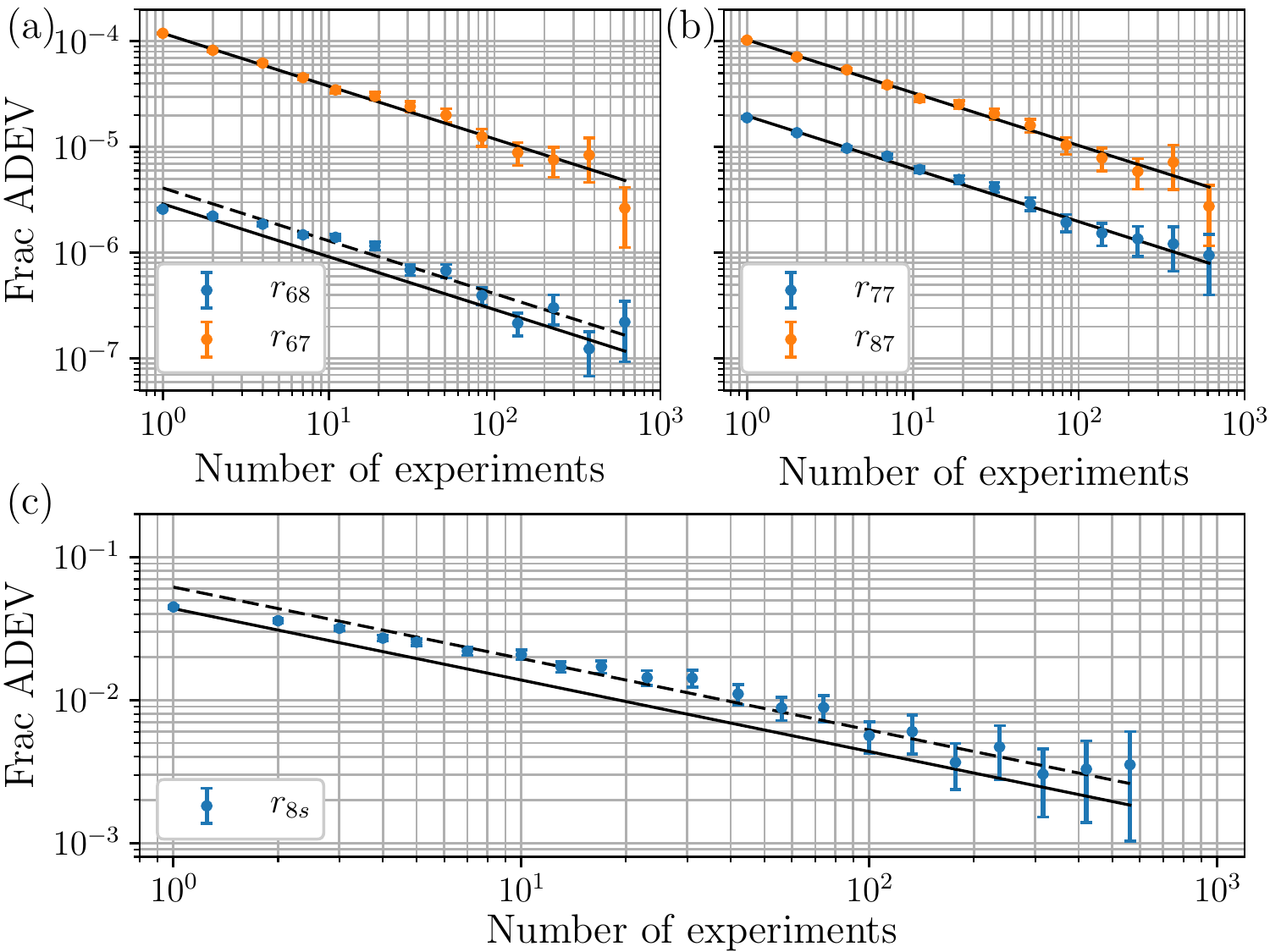}
\caption{ (a-c) Fractional Allan deviation of the ratios among $g_I$, $g_{F'}$.  Solid black lines indicate the projection noise limit, and dashed lines for a factor of $\sqrt{2}$ larger.}
\label{fig:r678}
\end{figure}

Other systematic effects considered include shifts on $^{176}$Lu$^+$ arising from the 1762-nm laser, the ac-magnetic field effect on Lu$^+$, and shifts on $^{138}$Ba$^+$ arising from microwave fields. These shifts are all well below the stated uncertainties and omitted from the table.
\begin{table}
\caption{Values and uncertainties for $r_6$ and $r_8$ including correction for systematic effects: $(xx)$ indicates the uncertainty for a given quantity, and $[-E]$ indicates a power of 10 $(\times 10^{-E})$.}
\label{tab:baratio}
\begin{ruledtabular}
\begin{tabular}{l r r}
Description & \multicolumn{1}{c}{Lu$^+$ left} &\multicolumn{1}{c}{Lu$^+$ right} \\\hline
\midrule
$r_6$ raw & 5.5965567(17) & 5.5965292(20) \\  \hline
magnetic gradient  & -1.3258(51)[-5] & 1.3258(51)[-5] \\
microwave ac Stark &7.41(69)[-7] &7.41(69)[-7] \\
ac magnetic field & -5.31(12)[-7] & -5.31(12)[-7]\\ \hline
$r_6$ corrected &5.5965437(17) &5.5965427(20)\\ 
\midrule \hline \hline \midrule
$r_8$ raw & -6.4770416(30) & -6.4770088(22) \\  \hline
magnetic gradient  & 1.5347(59)[-5] &-1.5347(59)[-5] \\
microwave ac Stark &6.24(43)[-7] & 6.24(43)[-7]\\
ac magnetic field & 6.15(14)[-7]& 6.15(14)[-7] \\ \hline
$r_8$ corrected & -6.4770250(30) & -6.4770229(22) \\ 
\end{tabular}
\end{ruledtabular}
\end{table}
After accounting for systematic effects, the ratio results $r_6$ and $r_8$ are in statistical agreement for both crystal configurations as seen in Table \ref{tab:baratio}.  For the final values, we take the weighted mean of results for the two crystal configurations.  As the measurements are not projection noise limited, we use the larger uncertainty from the two configurations in each case giving
\begin{subequations}
\label{Eq:rk}
\begin{align}
r_{6} =  & \frac{g_\mathrm{Ba}}{g_6} = 5.5965433(20), \label{Eq:r6} \\
r_{8} =  & \frac{g_\mathrm{Ba}}{g_{8}} = -6.4770236(30) \label{Eq:r8}.
\end{align}
\end{subequations}
Measurements on the single ion yields the following ratios,
\begin{subequations}
\label{Eq:g_ratio}
\begin{align}
r_{68} &= \frac{g_6}{g_8} = -1.157326607(88), \label{Eq:r68}\\
r_{67} &= \frac{g_6}{g_{7}} = -8.5026437(24),\\
r_{87} &= \frac{g_8}{g_{7}} = 7.3467970(20),\\
r_{77} &= \frac{g_{7}}{g_{7}^\prime} = 0.99999978(38),\\
r_{8s} &= \frac{g_8}{g_{I}} = -254.2897(17),
\end{align}
\end{subequations}
where $r_{77}$ is the ratio of the $\ket{^3D_1,7,\pm1}$ Zeeman splittings measured independently starting from either $\ket{^3D_1,6,0}$ or $\ket{^3D_1,8,0}$ and is statistically consistent with one as expected. The fractional Allan deviations are shown in \fref{fig:r678}(a-c). Again the statistical uncertainties of $r_{68}$ and $r_{8s}$ stated are given as $\sqrt{2}$ larger than the projection noise limited uncertainty as indicated by Allan deviations in \fref{fig:r678}(a,c). With microwave and optical interrogation times of $16\ms$ and $45\ms$, respectively, the systematic effects including shifts caused by the microwave fields and 848-nm light are negligible compared with stated statistical uncertainties. 

To check the consistency of the results, $r_{68}$ can be independently evaluated from {Eq.~(\ref{Eq:r6}-\ref{Eq:r8})} to give $r_{68} = \frac{r_8}{r_6} = -1.15732574(68)$, which can be compared with the directly measured value in \eref{Eq:r68}.  To determine $g_I$ and $g_F'$, we take the weighted mean of the two values $r_{68} = -1.157326593(88)$, and
\begin{align}
g_\mathrm{Ba} &= \frac{1}{2}\left[1.20036731(24)-2.00249492(3)\right]\nonumber\\
&=-0.40106232(12)
\end{align} 
determined from reported $g$-factors for $^{138}$Ba$^+$ \cite{marx1998precise,arnold2020precision}.  The values of $g_I$ and $g_{F'}$ are then determined to be:
\begin{subequations}
\label{Eq:g_factor}
\begin{align}
g_6 &=\frac{g_\mathrm{Ba}}{r_6} = -0.071662506(33),\\
g_7 &=\frac{g_\mathrm{Ba}}{r_6 r_{68} r_{87}} = 0.0084282619(46),\\
g_8 &=\frac{g_\mathrm{Ba}}{r_6 r_{68}} = 0.061920729(29),\\
g_I &=\frac{g_\mathrm{Ba}}{r_6 r_{68} r_{8s}} = -0.0002435047(16).
\end{align}
\end{subequations}
\section{Discussion}
From Appendix~\ref{sub:gfactor}, the $g$-factors for $^3D_1$ may be written
\begin{subequations}
\begin{align}
g_6&=-\tfrac{1}{7}g_J+\tfrac{8}{7}g_I\nonumber\\
&\quad+\tfrac{8}{35}\sum_{J'}\beta^1_{1,J'}-\tfrac{136}{455}\sum_{J'}\beta^2_{1,J'}+\delta g^{(2)}_6\\
g_7&=\tfrac{1}{56}g_J+\tfrac{55}{56}g_I\nonumber\\
&\quad+\tfrac{221}{840}\sum_{J'}\beta^1_{1,J'}-\tfrac{17}{280}\sum_{J'}\beta^2_{1,J'}+\delta g^{(2)}_7\\
g_8&=\tfrac{1}{8}g_J+\tfrac{7}{8}g_I\nonumber\\
&\quad+\tfrac{7}{40}\sum_{J'}\beta^1_{1,J'}+\tfrac{7}{40}\sum_{J'}\beta^2_{1,J'}+\delta g^{(2)}_8.
\end{align}
\end{subequations}
Neglecting $\delta g^{(2)}_7$ and using the measured values of $g_I$ and $g_F$, these equations can be solved for $g_J$, $\sum_{J'}\beta^1_{1,J'}$, and $\sum_{J'}\beta^2_{1,J'}$, which gives
\begin{subequations}
\begin{align}
\label{uncorrected}
g_J&=0.49823832(31),\\
\sum_{J'}\beta^1_{1,J'}&=-8.65774(41)\times 10^{-4}, \label{subEq:M1}\\
\sum_{J'}\beta^2_{1,J'}&=3.152(13)\times 10^{-5}, \label{subEq:Q2}
\end{align}
\end{subequations}
where uncertainties have been propagated from the measurements of $g_\mathrm{Ba}$, $r_6$, $r_{68}$, $r_{87}$, and $r_{8s}$.

To determine the corrections from $\delta g^{(2)}_F$, we first note that they can be expressed in terms of  $\beta^k_{1,2}$ and  $\redmel{^3D_2}{\mathbf{m}}{^3D_1}$.  Both $\sum_{J'}\beta^1_{1,J'}$ and $\sum_{J'}\beta^2_{1,J'}$ are largely determined by the $^3D_2$ contribution such that they can be used to approximate $\beta^k_{1,2}$ to better than $2\%$.  In addition, calculated matrix elements of $\mathbf{m}$ are typically accurate at the $1\%$ level.  Thus we evaluate Eq.~\ref{Eq:dg2} using $\redmel{^3D_1}{\mathbf{m}}{^3D_2}=-2.055\mu_B$ from \cite{paez2016atomic}, and Eq.~\ref{subEq:M1} and~\ref{subEq:Q2} for $\beta^1_{1,2}$ and $\beta^2_{1,2}$, respectively.  The resulting values of $\delta g_F^{(2)}$ are given in Eq.~\ref{Values:g2F}, which give the corrected values
\begin{subequations}
\label{corrected}
\begin{align}
g_J&=0.4982366(12),\\
\sum_{J'}\beta^1_{1,J'}&=-8.6574(29)\times 10^{-4},\\
\sum_{J'}\beta^2_{1,J'}&=3.130(21)\times 10^{-5}.
\end{align}
\end{subequations}
For $\sum_{J'}\beta^1_{1,J'}$ in particular, there is a significant cancellation of the corrections arising from those terms in Eq.~\ref{Eq:dg2} proportional to $g_F(J)-g_F(J')$ and those proportional to $W_{JF}-W_{J'F}$.  Consequently, it may well be that additional terms are needed to properly evaluate the corrections.  Instead, for all cases, we have used the largest of the two correction terms as the corresponding uncertainty when determining the overall uncertainties given in Eq.~\ref{corrected}.  The resulting values in Eq.~\ref{corrected}(b) and (c) are within 3\% and 17\%, respectively, of the theoretical estimates given in the Appendix, which is reasonable given the estimated uncertainties for calculated matrix elements given in \cite{paez2016atomic}.

Finally, the parameter $\beta^Q_{J,J'}$ from Eq.~\ref{Eq:quadP} associated with hyperfine-mediated quadrupole corrections may be written
\begin{equation}
\beta^Q_{J,J'}=\beta^1_{1,J'} \frac{\mu_B I \redmel{J'}{\Theta^{(2)}}{J}}{\redmel{J'}{\mathbf{m}}{J}}.
\end{equation}
Using the same approximations as above for $\beta^1_{1,2}$ and matrix elements in Appendix~\ref{ME}, gives $\beta^Q_{1,2}=-0.0133\, e a_0^2$.  Thus
\begin{equation}
\langle\delta \Theta(J,F,m)\rangle_F\approx\frac{2}{105}\beta^Q_{1,2}=-2.54\times 10^{-4}\,e a_0^2,
\end{equation}
As measured quadrupole moments are in agreement with theory to within $3\%$, we would expect the above estimate to be accurate to the 5\% level. This represents the effective quadrupole moment for the hyperfine averaged reference frequency for the $^1S_0-{}^3D_1$ clock transition.  For a $2\pi\times 200\,\mathrm{kHz}$ dc confinement, this would result in a maximum fractional frequency shift of $7 \times 10^{-19}$.  In practice, this would be suppressed by the field orientation technique demonstrated in \cite{tan2019suppressing}, which tunes the spatial dependence to zero leaving predominantly stray field contributions that may not be well aligned to the trap principle axes. 

In summary, we have carried out precision measurements of $g$-factors for the $^1S_0$ and $^3D_1$ levels of $^{176}$Lu$^+$.  These measurements provide direct evidence of hyperfine-mediated mixing for clock states in $^{176}$Lu$^+$, an accurate assessment of $g_J(^3D_1)$, and an estimate of a hyperfine-mediated quadrupole moment that is not cancelled by hyperfine-averaging.  Although the corresponding shift of the clock frequency will likely be well below $10^{-18}$ for typical operating conditions, it will inevitably be an important consideration for upcoming clock assessments for this atom.
\begin{acknowledgments}
We thank Sergey Porsev for identifying the correct sign dependencies between the reduced matrix elements.  This work is supported by the National Research Foundation, Prime Ministers Office, Singapore and the Ministry of Education, Singapore under the Research Centres of Excellence programme.  It is also supported by the National Research Foundation, Singapore, hosted by National University of Singapore under the Quantum Engineering Programme (Grant Award QEP-P5).  M.S.S. acknowledges the sponsorship of ONR Grants No. N00014-17-1-2252 and N00014-20-1-2513.
\end{acknowledgments}
\appendix
\section{Theory}
\label{Sect:Theory}
In this section relevant theoretical results for $g$-factor and quadrupole moments are given.  Explicit expressions are given for the $^3D_1$ states but results can be readily applied to $^3D_2$ and $^1D_2$.
\subsection{Hyperfine Interaction Theory}
From the relativistic treatment in \cite{beloy2008hyperfine}, the hyperfine Hamiltonian can be written as a sum of multipole interactions between electrons and nucleons,
\begin{equation}
H_{\mathrm{hfs}} = \sum_{k=1}^\infty\vb*{T}^e_k\cdot\vb*{T}^n_k,
\end{equation}
where  $\vb*{T}^e_{k}$ and $\vb*{T}^n_{k}$ are spherical tensor operators of rank $k$ that operate on the space of electronic and nuclear coordinates, respectively.  In the presence of the hyperfine interaction, the total angular momentum $\vb*{F} = \vb*{I}+\vb*{J}$ is conserved and basis states can be denoted $\ket{\gamma IJFm_F}$ where $\gamma$ denotes all other quantum numbers.  From the Wigner-Eckart theorem, a matrix element of $H_{\mathrm{hfs}}$ over the basis set is,
\begin{multline}
\matrixel{\gamma' IJ'F'm'}{H_{\mathrm{hfs}}}{\gamma IJFm}
= \delta_{FF'}\delta_{m',m}(-1)^{J'+I+F}\\
\times\sum\limits_{k=1}^{k'}\begin{Bmatrix}
F & J' & I \\
k & I & J
\end{Bmatrix}\redmel{\gamma' J'}{\vb*{T}^e_{k}}{\gamma J}\redmel{I}{\vb*{T}^n_{k}}{I},
\label{eq:melHFI}
\end{multline}
where $k'=\boldsymbol{\mathrm{min}}(2I,J+J')$.  For notational convenience we will drop $\gamma$ and $I$ in the notion.  As we are primarily concerned with the upper clock states, we will use $J=1,2,3,$ and $S$ to denote the triplet and singlet $D$ states. 

Following \cite{beloy2008hyperfine}, we will use the notation for the ``stretched" matrix element of a tensor operator $O_{k,q}$:
\begin{align}
\langle O_k\rangle_I&\equiv \langle II | O_{k,0}| II\rangle\nonumber\\
&=\begin{pmatrix}I & k & I\\-I & 0 & I\end{pmatrix} \redmel{I}{O_k}{I}.
\end{align}
In particular the nuclear magnetic dipole and electric quadrupole moments are defined as 
\begin{equation}
\mu_I=\langle \vb*{T}^n_{k} \rangle_I,\quad \mbox{and}\quad Q=2\langle \vb*{T}^n_{k} \rangle_I.
\end{equation}

For a given interaction, $H_I$, the first order energy shift $\mel{J F m}{H_I}{J F m}$ is modified by the hyperfine interaction.  Following the treatment of the quadrupole moment in \cite{beloy2017hyperfine}, the modification can be determined by treating $H_I$ and $H_\mathrm{hfs}$ on an equal footing in perturbation theory.  Explicitly, states are expanded to order $n$ in the hyperfine interaction and modification to the expectation value of $H_I$ is then attributed to a state dependent correction to the relevant property of the atom.  In considering the importance of various terms it should be noted that hyperfine interaction terms drop off significantly with $k$ such that $k=2$ terms at order $n$ can be comparable to $k=1$ terms at order $n+1$.

\subsection{Land\'{e} $\mathbf{g_F}$-factors}
\label{sub:gfactor}
With the Zeeman interaction
\begin{equation}
H_z=-\mathbf{m}\cdot\mathbf{B}=\frac{\mu_B B}{\hbar}\left(g_L L_z+g_S S_z+g_I I_z\right),
\end{equation}
the term from first order perturbation theory $\mel{J F m}{H_z}{J F m}$ is the usual weak field Zeeman shift $m g_F \mu_B B$.  Corrections derived from $n^{th}$-order perturbation theory in the hyperfine interaction also have a proportionality to $m \mu_B B$ and thus represent a correction to $g_F$, which we denote by $\delta g^{(n)}_F$.  Up to $n=1$ we have
\begin{multline}
\matrixel{\overline{JFm}}{H_z}{\overline{JFm}}=\matrixel{JFm}{H_z}{JFm}\\
+2\sum_{J'}\frac{\matrixel{JFm}{H_z}{J'Fm}\matrixel{J'Fm}{H_{\mathrm{hfs}}}{JFm}}{E_J-E_{J'}},
\end{multline}
from which we obtain
\begin{equation}
\delta g^{(1)}_F=2\sum_{J'}\frac{\matrixel{JFm}{H_z}{J'Fm}\matrixel{J'Fm}{H_{\mathrm{hfs}}}{JFm}}{m\mu_B B (E_J-E_{J'})}
\label{gFactor2}
\end{equation}
Since $H_z$ is a rank 1 tensor, only couplings to $J'=J\pm1$ contribute. Using 
\begin{multline}
\matrixel{JFm}{\mathbf{m}}{J'Fm}=\frac{m(-1)^{F+1+I+J'}(2F+1)}{\sqrt{F(F+1)(2F+1)}}\\
\times \begin{Bmatrix} F & F & 1\\ J & J' & I\end{Bmatrix}  \redmel{J}{\mathbf{m}}{J'},
\end{multline}
and noting that $I+J'+F$ must be integer, we have
\begin{multline}
\delta g^{(1)}_F=2\sqrt{\frac{2F+1}{F(F+1)}}\sum_{J',k}\begin{Bmatrix} F & F & 1\\ J & J' & I\end{Bmatrix}\begin{Bmatrix}F & J' & I \\k & I & J\end{Bmatrix} \\
\times\frac{\redmel{J}{\mathbf{m}}{J'}\redmel{J'}{\vb*{T}^e_{k}}{J}\redmel{I}{\vb*{T}^n_{k}}{I}}{\mu_B(E_J-E_{J'})},
\end{multline}
which may be written
\begin{equation}
\delta g^{(1)}_F=\sum_{k,J'}C^k_{F,J,J'}\beta^k_{J,J'},
\end{equation}
where
\begin{align}
\beta^1_{J,J'}&=\frac{\redmel{J}{\mathbf{m}}{J'}\redmel{J'}{\vb*{T}^e_{1}}{J}}{E_J-E_{J'}}\frac{\mu_I}{\mu_B I},\\
\beta^2_{J,J'}&=\frac{\redmel{J}{\mathbf{m}}{J'}\redmel{J'}{\vb*{T}^e_{2}}{J}}{E_J-E_{J'}}\frac{Q}{2\mu_B I},
\end{align}
and
\begin{multline}
C_{F,J,J'}^k=2I\sqrt{\frac{(2F+1)}{F(F+1)}}\begin{pmatrix}I & k & I\\-I & 0 & I\end{pmatrix}^{-1}\\
\times\begin{Bmatrix} F & F & 1\\ J & J' & I\end{Bmatrix} \begin{Bmatrix}F & J' & I \\k & I & J\end{Bmatrix},
\end{multline}

For $^3D_1$, we have
\begin{subequations}
\begin{align}
\delta g^{(1)}_6&=\tfrac{8}{35}\sum_{J'}\beta^1_{1,J'}-\tfrac{136}{455}\sum_{J'}\beta^2_{1,J'}\\
\delta g^{(1)}_7&=\tfrac{221}{840}\sum_{J'}\beta^1_{1,J'}-\tfrac{17}{280}\sum_{J'}\beta^2_{1,J'}\\
\delta g^{(1)}_8&=\tfrac{7}{40}\sum_{J'}\beta^1_{1,J'}+\tfrac{7}{40}\sum_{J'}\beta^2_{1,J'}.
\end{align}
\end{subequations}
The dominant contribution is from the M1 coupling to $^3D_2$, for which $\beta^1_{1,2} = -9.1\times 10^{-4}$, using matrix elements given in Appendix~\ref{ME}.  The contribution from $^1D_2$ is less than $2\%$ of that with $\beta^1_{1,S}=1.5\times 10^{-5}$ and the $k=2$ corrections from $^3D_2$ have a similar magnitude with $\beta^2_{1,2}=2.6\times 10^{-5}$.  At this few percent level, one should consider next order corrections, given by
\begin{widetext}
\begin{multline}
m\mu_B \delta g^{(2)}_F=-\sum_{J'\neq J}\sum_{J''\neq J}\Bigg[\frac{2\mel{JFm}{\mathbf{m}}{J''Fm}\mel{J''Fm}{H_\mathrm{hfs}}{J'Fm}\mel{J'Fm}{H_\mathrm{hfs}}{JFm}}{(E_{J}-E_{J''})(E_{J}-E_{J'})}\\
+\frac{\mel{JFm}{H_\mathrm{hfs}}{J''Fm}\mel{J''Fm}{\mathbf{m}}{J'Fm}\mel{J'Fm}{H_\mathrm{hfs}}{JFm}}{(E_{J}-E_{J''})(E_{J}-E_{J'})}\Bigg]+\mel{JFm}{\mathbf{m}}{JFm}\sum_{J'\neq J}\frac{|\mel{JFm}{H_\mathrm{hfs}}{J'Fm}|^2}{(E_{J}-E_{J'})^2}\\
+2\mel{JFm}{H_\mathrm{hfs}}{JFm}\sum_{J'\neq J}\frac{\mel{JFm}{\mathbf{m}}{J'Fm}\mel{J'Fm}{H_\mathrm{hfs}}{JFm}}{(E_{J}-E_{J'})^2}.
\end{multline}
For $J=1$, this is dominated by coupling to $^3D_2$ for which $J'=J''=2$.  Hence
\begin{multline}
\delta g^{(2)}_F\approx \left(g_F(J')-g_F(J)\right)\frac{|\mel{JFm}{H_\mathrm{hfs}}{J'Fm}|^2}{(E_{J}-E_{J'})^2}+\left(W_{JF}-W_{J'F}\right)\frac{2\mel{JFm}{\mathbf{m}}{J'Fm}\mel{J'Fm}{H_\mathrm{hfs}}{JFm}}{(E_{J}-E_{J'})^2},
\end{multline}
where $W_{JF}=\mel{JFm}{H_\mathrm{hfs}}{JFm}$ are the diagonal matrix elements of the hyperfine interaction.  Taking only the $k=1,2$ contributions for the off-diagonal matrix elements and using the definitions of $\beta_{1,2}^k$, we have
\begin{equation}
\label{Eq:dg2}
\delta g^{(2)}_F\approx \left(g_F(J')-g_F(J)\right)\left|\sum_{k=1}^2\begin{Bmatrix}F & J' & I \\ k & I & J\end{Bmatrix}\begin{pmatrix}I & k & I\\-I & 0 & I\end{pmatrix}^{-1}\frac{\mu_B I \beta^k_{J,J'}}{\redmel{J}{\mathbf{m}}{J'}}\right|^2
-\frac{W_{JF}-W_{J'F}}{E_{J}-E_{J'}}\sum_{k=1}^2 C^k_{F,J,J'}\beta^k_{J,J'},
\end{equation}
\end{widetext}  
Following \cite{kaewuam2019spectroscopy}, $W_{JF}$ can be expressed in terms of the measured hyperfine splittings \cite{kaewuam2017laser,kaewuam2019spectroscopy} and a smaller hyperfine-induced scalar shift common to all $F$ levels of a given $J$.  Neglecting the scalar contributions, we obtain the estimates
\begin{subequations}
\label{Values:g2F}
\begin{align}
\delta g^{(2)}_6&\approx -3.29\times 10^{-7},\\
\delta g^{(2)}_7&\approx 1.10\times 10^{-8},\\
\delta g^{(2)}_8&\approx 2.65\times 10^{-7},
\end{align}
\end{subequations}
where we have approximated $g_F$ using $g_J=1/2$ and neglected $g_I$.  Hence, $|\delta g^{(2)}_F|\lesssim 1\times 10^{-3}g_I$.  
\subsection{Quadrupole moments}
A similar treatment can be applied to determine hyperfine-mediated quadrupole moments.  In this case the resulting quadrupole correction does not average to zero and will thus be a limitation to hyperfine averaging.  The analogous expression for the quadrupole correction is 
\begin{multline}
\matrixel{\overline{JFm}}{H_Q}{\overline{JFm}}=\matrixel{JFm}{H_Q}{JFm}\\
+2\sum_{J'}\frac{\matrixel{JFm}{H_Q}{J'Fm}\matrixel{J'Fm}{H_{\mathrm{hfs}}}{JFm}}{E_J-E_{J'}}.
\label{HFCQuadrupoel}
\end{multline}
The first term in this expression is exactly as derived by Itano \cite{ItanoQuad} and can be written
\begin{equation}
\label{QuadShift}
\matrixel{JFm}{H_Q}{JFm}=C_{F,m}\Theta(J) f(\alpha,\beta),
\end{equation}
where
\begin{multline}
C_{F,m}= (-1)^{2F+I+J-m}(2F+1)\\
\times \begin{pmatrix} F & 2 & F\\ -m & 0 & m\end{pmatrix}\begin{Bmatrix} F & F & 2\\ J & J & I\end{Bmatrix}\begin{pmatrix} J & 2 & J\\ -J & 0 & J\end{pmatrix}^{-1},
\end{multline}
$\Theta(J)$ is the usual quadrupole moment for the fine-structure level defined by
\begin{equation}
\Theta(J)=\begin{pmatrix} J & 2 & J\\ -J & 0 & J\end{pmatrix}\redmel{J}{\Theta^{(2)}}{J},
\end{equation}
and $f(\alpha,\beta)$ is determined by the orientation and strength of the applied external field.  With the potential in the principle axis frame given by
\begin{equation}
\phi=A\left[x^2+y^2-2 z^2+\epsilon\left(x^2-y^2\right)\right]
\end{equation}
we have
\begin{multline}
f(\alpha,\beta)=-A\big[(3\cos^2\beta-1)\\
-\epsilon\sin^2\beta(\cos^2\alpha-\sin^2\alpha)\big]
\end{multline}
where $\alpha$ and $\beta$ are the Euler angles as defined in \cite{ItanoQuad}.  

The matrix element $\matrixel{JFm}{H_Q}{J'Fm}$ can be found in the same way as Eq.~\ref{QuadShift} giving
\begin{multline}
\matrixel{JFm}{H_Q}{J'Fm}=(-1)^{2F+I+J'-m}(2F+1)\\
\times \begin{pmatrix} F & 2 & F\\ -m & 0 & m\end{pmatrix}\begin{Bmatrix} F & F & 2\\ J & J' & I\end{Bmatrix} \redmel{J}{\Theta^{(2)}}{J'} f(\alpha,\beta).
\end{multline}
As it has the same orientation dependence as Eq.~\ref{QuadShift}, the correction can be viewed as a change in the state-dependent quadrupole moment $\Theta(J,F,m)=C_{F,m} \Theta(J)$ by $\delta\Theta(J,F,m)$, which may be written
\begin{widetext}
\begin{equation}
\delta\Theta(J,F,m)=2(2F+1)(-1)^{F-m}\begin{pmatrix} F & 2 & F\\ -m & 0 & m\end{pmatrix}
\sum_{J',k} \begin{Bmatrix} F & F & 2\\ J & J' & I\end{Bmatrix} \begin{Bmatrix}F & J' & I \\k & I & J\end{Bmatrix}\frac{\redmel{J}{\Theta^{(2)}}{J'}\redmel{J'}{\vb*{T}^e_{k}}{J}\redmel{I}{\vb*{T}^n_{k}}{I}}{E_J-E_{J'}}.
\end{equation}
Taking only the $k=1$ terms gives
\begin{equation}
\delta\Theta(J,F,m)=2(2F+1)(-1)^{F-m}
\begin{pmatrix} F & 2 & F\\ -m & 0 & m\end{pmatrix} \begin{pmatrix} I & 1 & I\\ -I & 0 & I\end{pmatrix}^{-1}\sum_{J'} \begin{Bmatrix} F & F & 2\\ J & J' & I\end{Bmatrix} \begin{Bmatrix}F & J' & I \\1 & I & J\end{Bmatrix}\beta^Q_{J,J'},
\label{Eq:quad}
\end{equation}
\end{widetext}
where 
\begin{equation}
\label{Eq:quadP}
\beta^Q_{J,J'}=\frac{\redmel{J}{\Theta^{(2)}}{J'}\redmel{J'}{\vb*{T}^e_{1}}{J}}{E_J-E_{J'}}\mu_I.
\end{equation}
For $^3D_1$, the only contributions are from $^3D_2$ and $^1D_2$.  For the $m=0$ states of interest
\begin{subequations}
\begin{align}
\delta \Theta(J,6,0)&= -\tfrac{16}{175}\sum_{J'}\beta^Q_{1,J'}\\
\delta \Theta(J,7,0)&=\tfrac{1}{35}\sum_{J'}\beta^Q_{1,J'}\\
\delta \Theta(J,8,0)&=\tfrac{3}{25}\sum_{J'}\beta^Q_{1,J'}.
\end{align}
\end{subequations}
The average over $F$ is given by
\begin{equation}
\langle\delta \Theta(J,F,m)\rangle=\frac{2}{105}\sum_{J'}\beta^Q_{1,J'},
\end{equation}
which is independent of $m$ at this level of approximation.  The dominant term is again the $^3D_2$ contribution for which $\beta^Q_{1,2}=-0.014$.  Omitting the $^1D_2$ contribution, we get a theoretical estimate of $-2.63\times10^{-4} e a_0^2$ for the effective quadrupole moment of the hyperfine-averaged transition.
\subsection{Matrix Elements}
\label{ME}
Matrix elements used in this work are from results reported in Ref.~\cite{paez2016atomic,kaewuam2019spectroscopy}.  However, signs of matrix elements are not always specified, as the sign of a single matrix element can be set arbitrarily.  As this work explicitly requires the relative sign between matrix elements, we give a list of the relevant matrix elements inclusive of sign in table~\ref{MatrixElements}.  Matrix elements of $\vb*{T}^e_{2}$ given in the table differ in sign from those given in \cite{kaewuam2019spectroscopy}.  This was due to a difference in the definition of $\vb*{T}^e_{2}$ relative to \cite{beloy2008hyperfine} that was discovered in the course of this work.  This will result in minor changes to the calculated results in \cite{paez2016atomic,kaewuam2019spectroscopy} but not significantly influence the results or conclusions in those reports.
\begin{table}[h!]
\caption{\label{MatrixElements} Reduced matrix elements used in this work.  These are derived from the work in Ref.~\cite{paez2016atomic} and include the relative sign.}
\begin{ruledtabular}
\begin{tabular}{c c c c}
\toprule
ME & Value & ME & Value  \\
\colrule
$\langle {}^3D_2\|\,\vb*{T}^e_{1}\| {}^3D_1 \rangle$  &  -18682 & $\langle {}^1D_2\|\,\vb*{T}^e_{1}\| {}^3D_1 \rangle$  &  10618 \\
$\langle {}^3D_2 \|\,\vb*{T}^e_{2}\| {}^3D_1 \rangle$  &  686 & $\langle {}^1D_2 \|\,\vb*{T}^e_{2}\| {}^3D_1 \rangle$  &  70 \\
$\langle {}^3D_1\|\,\mathbf{m}\| {}^3D_2 \rangle$  &  -2.055 & $\langle {}^3D_1\|\,\mathbf{m}\| {}^1D_2 \rangle$  &  -0.524 \\
$\langle {}^3D_1 \|\,\Theta^{(2)}\| {}^3D_2 \rangle$  &  -4.523 & $\langle {}^3D_1 \|\,\Theta^{(2)}\| {}^1D_2\rangle$  &  -1.018 \\
\end{tabular}
\end{ruledtabular}
\end{table}
\bibliographystyle{unsrt}
\bibliography{gfactor}
\end{document}